# How do Machine Learning Projects use Continuous Integration Practices? An Empirical Study on GitHub Actions

João Helis Bernardo
Federal University of Rio Grande do Norte
Federal Institute of Rio Grande do Norte
Natal, Brazil
joao.helis@ifrn.edu.br

Daniel Alencar da Costa
University of Otago
Dunedin, New Zealand
danielcalencar@otago.ac.nz

Sérgio Queiroz de Medeiros
Federal University of Rio Grande do Norte
Natal, Brazil
sergio.medeiros@ufrn.br

Uirá Kulesza
Federal University of Rio Grande do Norte
Natal, Brazil
uira@dimap.ufrn.br

## ABSTRACT

Continuous Integration (CI) is a well-established practice in traditional software development, but its nuances in the domain of Machine Learning (ML) projects remain relatively unexplored. Given the distinctive nature of ML development, understanding how CI practices are adopted in this context is crucial for tailoring effective approaches. In this study, we conduct a comprehensive analysis of 185 open-source projects on GitHub (93 ML and 92 non-ML projects). Our investigation comprises both quantitative and qualitative dimensions, aiming to uncover differences in CI adoption between ML and non-ML projects. Our findings indicate that ML projects often require longer build duration, and medium-sized ML projects exhibit lower test coverage compared to non-ML projects. Moreover, small and medium-sized ML projects show a higher prevalence of increasing build duration trends compared to their non-ML counterparts. Additionally, our qualitative analysis illuminates the discussions around CI in both ML and non-ML projects, encompassing themes like *CI Build Execution and Status*, *CI Testing*, and *CI Infrastructure*. These insights shed light on the unique challenges faced by ML projects in adopting CI practices effectively.

## CCS CONCEPTS

• **Computing methodologies → Machine learning**.

## KEYWORDS

continuous integration, machine learning, github actions, mining software repositories

## 1 INTRODUCTION

The recent trend in the software industry involves integrating artificial intelligence (AI) capabilities based on advances in machine learning (ML) techniques [1]. ML has been applied to various domains [2–4], gaining further prominence due to recent advancements, such as the ability to build Large Language Models (LLMs) [5]. It is estimated that over 50% of organizations are either exploring or in the planning stages of adopting ML technology [6]. ML techniques rely on both mathematics and software engineering [2]. While mathematics is used to generate the statistical models that are the basis of ML algorithms, software engineering is employed for the implementation and robust performance of the software project.

The AI domain has aspects fundamentally different from prior software application domains [1]. ML projects follow a distinct development process compared to traditional software projects, involving data engineering and model management [7], which encompass more intense data treatment and testing iterations. For instance, managing and versioning the data needed for ML applications is much more difficult than other types of software projects. Additionally, the customization and reuse of ML models demand specialized skills not commonly found within standard software teams [1]. Therefore, the research area of Software Engineering for Machine Learning (SE4ML) has employed a significant effort to understand and adapt software engineering practices to effectively develop, deploy, and maintain ML projects. ML projects also require multiple iterations to integrate new functionalities and improve their quality, and thus may benefit from Agile Releasing Engineering practices [8], such as Continuous Integration (CI) [9].

CI is a widely adopted practice that advocates frequent and automated integration of code changes into a shared repository, usually on a daily basis [10, 11]. Its principles emphasize not only frequent code integration but also automated testing, enabling rapid feedback loops and facilitating collaboration among development teams [12]. Prior research has investigated the impact of CI on the context of open-source projects [13–17]. For instance, Vasilescu et al. [13] revealed a positive association between the adoption of Travis CI and the number of bugs detected by core developers.

Hilton et al. [16] observed that CI is pivotal in facilitating more frequent software releases. Furthermore, recent works into CI have expanded their focus beyond the mere adoption of CI services, instead investigating the specific CI practices employed by software projects [18–20]. CI practices refer to common practices that should be followed by developers when adopting CI, such as frequent code commits, short build duration, quick build fixes, and test coverage monitoring. In the context of ML projects, Rzig et al. [9] conducted an analysis revealing that approximately 37% of ML projects have integrated a CI service in their workflow. They also found that the most prevalent CI tasks in ML projects encompass software testing and building. Furthermore, recent efforts aimed to create dedicated CI services for the ML field [21], considering its intricacies (i.e., traditional testing may overfit ML models).

Nevertheless, there is a knowledge gap concerning the adoption patterns of CI practices within ML projects. Questions such as "Do ML projects effectively employ CI practices?" have yet to be answered. Uncovering the nuances of CI adoption in ML projects can guide the development of customized approaches for the effective adoption of CI practices within this unique domain. Therefore, we propose an empirical study that analyzes data from 185 open-source projects from GitHub, comprising 93 ML projects and 92 non-ML projects. We aim to investigate how CI is applied and discussed by software developers in the ML development domain. Our investigation focuses on the following research questions (RQs):

- **RQ1. To what extent do CI practices adoption differ between ML and non-ML projects?** ML projects tend to require a longer build duration. In addition, medium-sized ML projects tend to have a lower test coverage.
- **RQ2. What are the evolution trends of build duration and test coverage within ML and non-ML projects?** Small and medium-sized ML projects manifest higher increasing build duration trends (75% and 61.4%) compared to non-ML projects (35.7% and 44.7%). Furthermore, both ML and non-ML projects manifest a maintaining test coverage trend, even with 46% of the medium ML projects exhibiting a median coverage rate below 75%.
- **RQ3. What do ML and non-ML developers discuss about CI in their projects?** Both ML and non-ML projects share common discussions on *CI Build Execution and Status*, *CI Infrastructure*, *CI Pipeline Configuration*, and *CI Testing and Code Quality*. However, ML projects exhibit a more extensive range of CI-related themes (73 themes) compared to non-ML projects (23 themes). Notably, a significant difference arises in the prevalence of the "relatedness of failures" theme in ML project discussions, indicating a potential higher incidence of false positives in their CI systems.

***Paper organization.*** In Section 2, we present our study setup, followed by the presentation of results and discussions in Sections 3 and 4, respectively. Section 5 addresses the threats to validity, while Section 6 discusses the related work. We conclude our paper in Section 7.

## 2 RESEARCH METHODOLOGY

In this section, we explain how we select the studied projects and construct the database that we use in our analyses.

### 2.1 Studied Projects

To fulfill our goal of studying CI practices in both ML and non-ML projects, we started selecting the dataset proposed by Gonzalez et al. [4] which was revised by Rzig et al. [9]. This is a curated dataset of 4,031 ML projects and 4,076 non-ML projects hosted on GitHub . The ML projects category comprises ML frameworks and libraries such as scikit-learn, and ML applications such as Faceswap. These projects leverage ML techniques or components and serve specific user needs or have general-purpose utility. In contrast, the non-ML projects category is composed of traditional software applications, including websites, desktop or mobile applications, and more. Notably, these non-ML projects are characterized by their absence of ML-based components in both design and functionality. We then conducted a comprehensive analysis by applying a series of filters to the initial dataset provided by Rzig et al. [9]. The selection process, as shown in Figure 1, involved the following steps:

***Step 1: Selecting projects updated within 2023 with default branch as 'main' or 'master'.*** We began the process by using the GitHub API to identify projects updated within 2023. Projects with a default branch different than 'main' or 'master' (e.g., 'dev', 'development', or 'draft') were excluded due to them not representing a user-intended project. The search was conducted using the GitHub API in June 2023, resulting in 6,915 projects (3,082 ML and 3,833 non-ML).

***Step 2: Selecting projects with GitHub Actions workflow configuration files.*** To select projects using GitHub Actions , we excluded projects lacking at least one *workflow* configuration file. This is a YAML file stored within the *'.github/workflows/'* folder of a GitHub repository, which defines a series of automated steps to be executed when specific events occur in the repository (i.e., push or pull request). This reduced the number of projects to 1,053 (553 ML and 500 non-ML).

***Step 3: Selecting projects with active GitHub Actions CI Workflows.*** Among the remaining projects, our next step was to identify those with at least one active GitHub Actions *CI workflow*. A repository can have multiple *Workflows*. However, unlike conventional CI/CD services, GitHub Actions *Workflows* can be used not only for executing test suites or deploying new releases (i.e., CI-related practices), but also to facilitate other software development activities, such as communication, code review, and dependency management [22]. Each *Workflow* may serve different purposes, for instance, one might handle CI tasks, such as compiling the codebase and executing the test suite, while another might perform tasks that extend beyond the scope of CI, like welcoming new contributors with their first pull request or even generating documentation automatically from code comments. In this study, our analysis centers on projects that have GitHub Actions *workflows* associated with CI, which we call *CI workflows*. For *workflows* to be classified as *CI workflows*, they must meet the following criteria: **(i) Exclusion of documentation references:** The workflow name or filename should not contain any documentation-related words, such as "docs.yml" or "documentation.yaml"; **(ii) Inclusion of code change event triggers:** The workflow should be triggered by either a pull_request or push event; **(iii) Inclusion of CI-related terminology:** The *workflow* configuration file must incorporate a CI-related word in either the *job* or *step* name. CI-related terms encompass various spellings, including "continuous integration," "continuousintegration," "continuous-integration," "continuous_integration," or simply "test." This inclusivity acknowledges that a fundamental task for any *CI workflow* is the execution of the test suite.

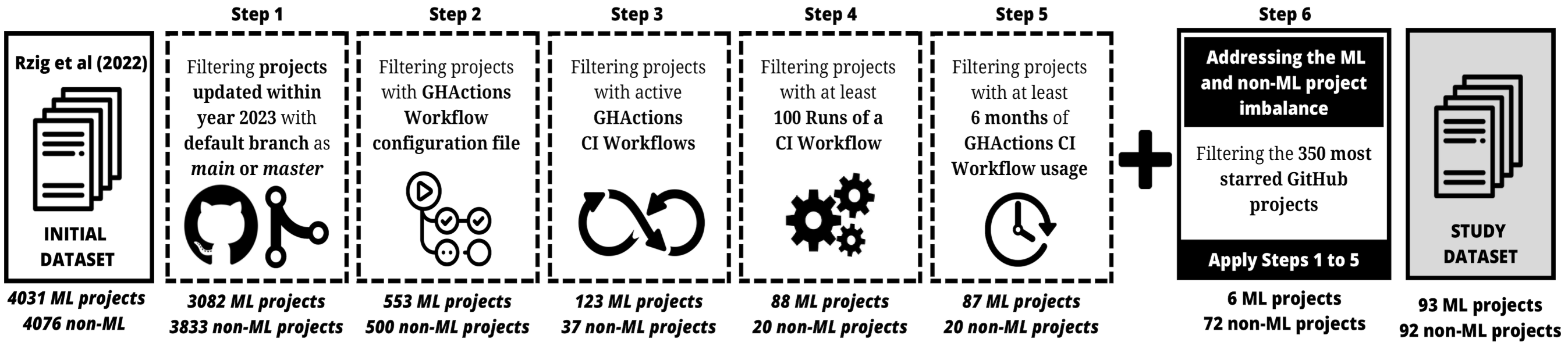

**Figure 1: An overview of the project selection process.**

To verify whether our workflow classification met our goals, we manually inspected a sample of 100 workflows from 74 projects of varying sizes. The inspection aimed to validate the criteria for automatically classifying the workflows as CI or non-CI. The first author manually inspected the workflow sample by reading their configuration files and assessed their content against the criteria. The results were validated with the third and fourth authors. Our criteria achieve 86.7% recall and 100% precision. Since there was no instance where a non-CI workflow was misclassified as a CI workflow, we leverage our criteria to select only projects that have CI workflows. 160 projects remained (123 ML and 37 non-ML).

***Step 4: Applying a threshold of at least 100 runs for CI workflows.*** We further filtered out projects that had fewer than 100 *Runs* in their GitHub Actions CI workflows. This threshold, which was inspired by previous studies [13, 15], seeks to ensure a significant volume of data for in-depth analysis. The adoption of a similar threshold in these studies reinforced our choice. 108 projects remained after this step (88 ML and 20 non-ML projects).

***Step 5: Selecting a minimum 6-month GitHub Actions CI workflow history.*** In order to ensure a substantial volume of monthly data intervals, we excluded projects containing less than 6 months of GitHub Actions CI Workflow historical data. We computed this criterion by calculating the period between the first and last *Run* of the GitHub Actions CI workflow for each project. 107 projects remained after this step (87 ML and 20 non-ML projects).

***Step 6: Addressing the ML and non-ML project imbalance.*** Given that we observed an imbalance between ML and non-ML projects during our project selection process, and to address this imbalance, we conducted an additional search on the GitHub API. We used an approach similar to Gonzalez et al. [4], selecting top projects based on stars (initially top 100, then increasing by 50). On these projects, we applied the filtering criteria of a minimum of 100 runs, 5 stars or forks, and updates in 2023. By adhering to these criteria, we successfully curated a well-balanced dataset after filtering the top 350 projects. These criteria are aligned with best practices for filtering out inactive or non-software repositories [23, 24]. Subsequently, the first author manually reviewed each project's repository page to classify it as ML or non-ML and exclude unsuitable projects (e.g., toy projects or tutorials). We identified and removed 9 projects that were not suitable for our analysis. For instance, the repository *"public-apis/public-apis"* is a collective list of free APIs for use in software and web development, which is not a real software development project. In contrast, despite the option to exclude ML projects, we chose inclusivity as we found influential ML projects like *"huggingface/transformers"* and *"pandas-dev/pandas"*, a robust data analysis library for Python [25].

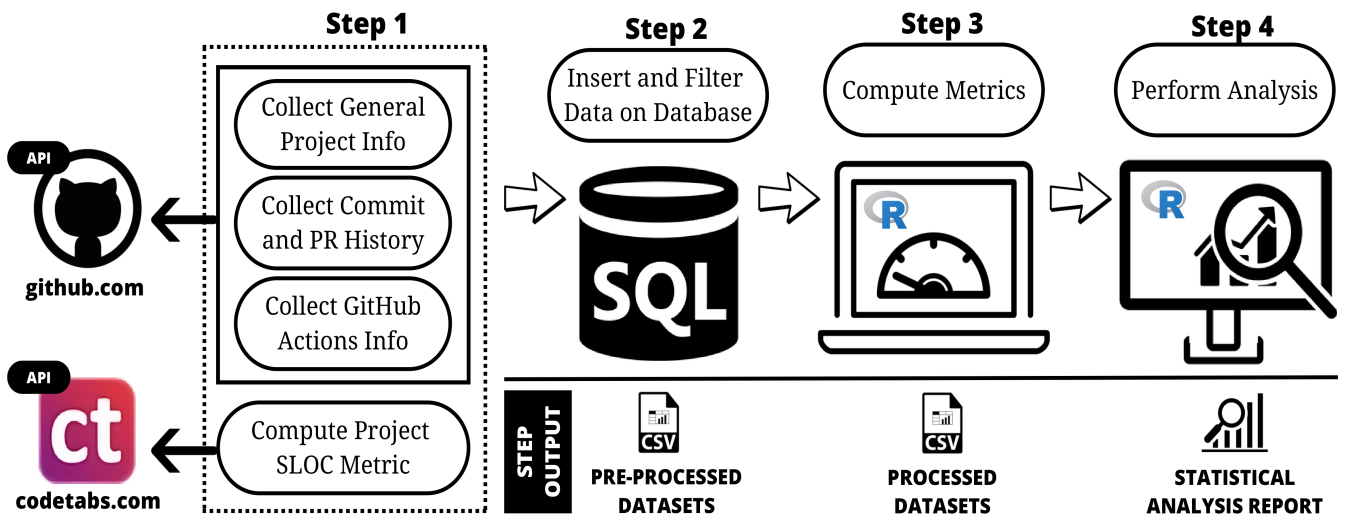

**Figure 2: An overview of the data collection process.**

We then applied the same filtering criteria used in the initial dataset (Steps 1 to 5) to these new projects. The criteria resulted in 78 projects (6 ML and 72 non-ML). The additional non-ML projects span diverse categories, including Static Site Generators, Cloud Tools, Web/Mobile Frameworks, package management, etc. As such, we consider that the non-ML projects are comparable with our set of ML projects in terms of complexity. Furthermore, we mitigate the effects of potential confounds in our analysis, for example, choosing a similar amount of small, medium, and large-sized projects for both ML and non-ML projects. Combined with the initial 107 projects filtered from the work of Rzig et al. [9], our final dataset contains 185 projects (93 ML and 92 non-ML). This balanced dataset allows us to analyze CI practices in both ML and non-ML projects.

## 2.2 Data Collection

After the selection of our studied projects, we begun the process of collecting project metadata. The data collection process is shown in Figure 2. Each step of the process is detailed below.

***Step 1: Collection of projects metadata.*** We use the GitHub API to collect general information for each studied project, such as primary language, size, and number of stars and forks, and to collect metadata related to the usage of GitHub Actions within these projects. For instance, we systematically compiled all active GitHub Actions workflows for the selected projects. Additionally, we retrieved metadata associated with each individual *Workflow Run* within CI workflows that have over 100 *Workflow Runs*. This selective approach concentrated solely on *completed Workflow Runs* triggered by *push* or *pull_request* events, thereby excluding *Workflow Runs* related to activities like pull request comments. This is because we intended to only analyze *Workflow Runs* associated with code changes within the projects. Finally, we used the GitHub API to collect all the commit and PR history of the projects within their default branch to the period after the adoption of GitHub Actions.

Moreover, we use the CodeTabs API[1] to collect the LOC (Lines of Code) metric for the analyzed projects. This API incorporates the `boyter/scc`[2] library, a robust tool for precise LOC counting. The CodeTabs API retrieves the LOC metric from any public GitHub repository with a size not greater than 500 Mb. A total of 8 of the studied projects exceeded this limit, then we manually used the *boyter/scc* library to calculate the LOC size of such projects. We use the LOC to categorize our projects into three size groups: small ($LOC <= 10,000$), medium ($10,000 < LOC <= 100,000$), and large projects ($LOC > 100,000$). We use these groups to perform comparisons in our analyses, which align with the methodology used by Felidre et al. [18]. In Table 1, we provide the general information (e.g., age) of the projects and the totals of analyzed commits, pull requests, coverage builds, and CI builds.

***Step 2: Dataset Preprocessing.*** After the collection of the project's metadata, we created a MySQL schema to systematically store and organize information related to entities such as GitHub repositories, GitHub Actions workflows, workflow runs, pull requests, and commits. With the project's metadata securely stored within the database, we used SQL queries to create pre-processed CSV datasets. These datasets were crafted specifically to include only the essential metadata critical to our subsequent analyses.

***Step 3: Compute Metrics.*** We use data from Steps 1 and 2 to compute the four CI metrics used in our analyses. To calculate the metrics, we segment the project history into periods of 30 days (one month), see Figure 3. In our analysis, we only consider projects containing at least 6 monthly intervals of analysis where GitHub Actions CI workflows are consistently used. Additionally, we exclude from our analysis the data related to the period of 15 days after the first CI workflow run, as it is already reported in the literature as an unstable period of adaptation of the usage of a CI service [14]. The formulas used to calculate the metrics associated with each CI practice were inspired by the work of Santos et al. [19] and Felidré et al. [18], and were collected from the GitHub, Coveralls [26], and CodeCov [27] APIs, which are web-based services that provide code coverage metrics for software projects. The description of each metric is presented in the following along with its data source (highlighted within brackets).

- ***Commit Activity*** [`GitHub`]: This metric represents the rate of days in the monthly interval that had at least one commit [19]. It ranges from 0 to 1, with 0 indicating that 0 commits were performed throughout the days of the time period, whereas 1 indicates that at least one commit was performed in each day of the time period.
- ***Build Duration*** [`GitHub`]: It quantifies the median duration of CI workflow runs in minutes, computed as the difference between the *run updated at timestamp* and the *run started at timestamp*. The duration is computed exclusively for successfully completed CI workflow runs. Failed workflow runs are excluded from this computation as they usually have a shorter build duration, stopping when the failure occurs.
- ***Time to Fix Broken Build*** [`GitHub`]: This metric represents the median time in hours that broken CI workflow runs remain in a broken state within the monthly interval. When a CI workflow run breaks, we compute the duration in hours until the workflow run returns to the "success" state. If a broken workflow run does not revert to the "success" state by the end of the monthly period, then this run is excluded from our analysis.
- ***Test Coverage*** [`Coveralls/CodeCov`]: It quantifies the percentage of code in a project that is covered by automated tests. This metric is collected from the last available build of a project in the Coveralls or CodeCov platforms. It is measured on a scale from 0 to 100, where 0 indicates that no line of code is covered by tests, and 100 indicates that every line of code is exercised by tests.

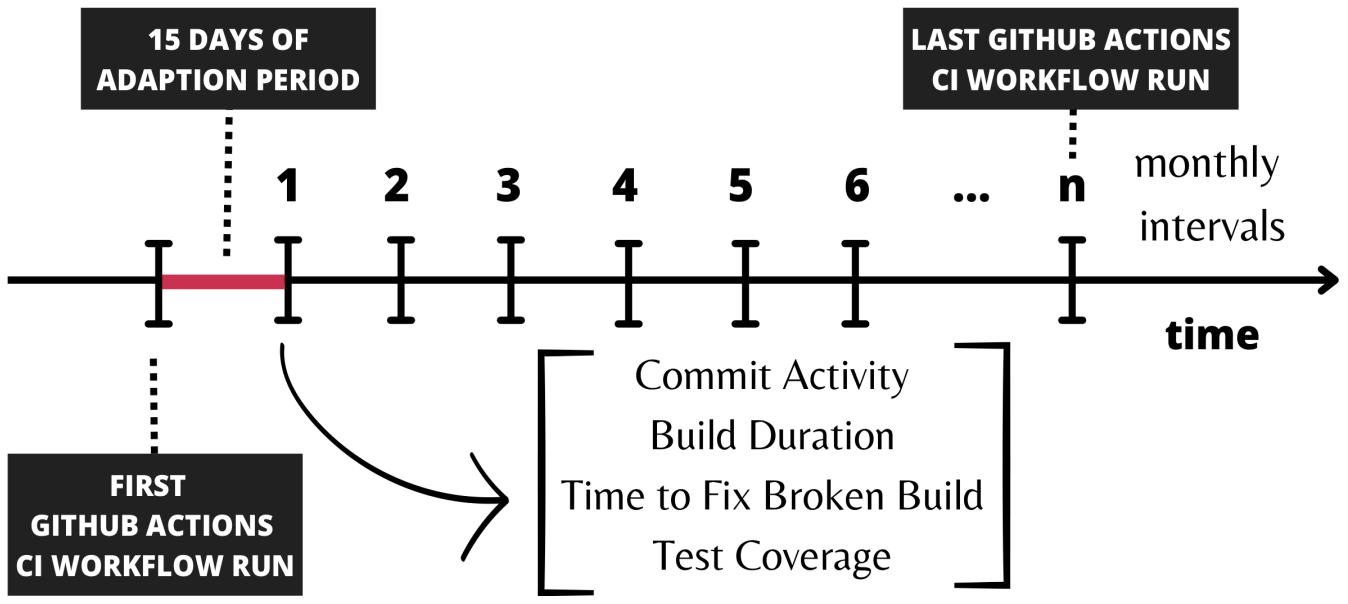

**Figure 3: Periods of Analysis.**

***Step 4: Perform Analysis.*** The curated datasets produced in Step 3 are the basis for our subsequent quantitative and qualitative analyses. We use the generated CSV datasets that incorporate computed CI practices' metrics, which are used as inputs for our statistical analyses. The analyses are performed using `R`. All the scripts and datasets we used in our analyses are available in our online Appendix[3] to the interested reader.

## 2.3 Research Questions Design

### RQ1. To what extent do CI practices adoption differ between ML and non-ML projects?

***Motivation.*** Adopting CI is beyond the sole implementation of a CI service, requiring effective use of recommended CI practices [12, 18, 28]. As such, the intricate nature of ML, involving complex algorithms and data, might present unique challenges in adopting CI practices effectively. For example, it might be harder to maintain shorter build duration in ML projects because they require more complex tests. Given the significant knowledge gap regarding the specific application of CI practices within ML projects, we start our investigations by comparing CI metrics from ML and non-ML projects.

[1]https://codetabs.com
[2]https://github.com/boyter/scc
[3]https://ci-ml-msr.github.io - Reproduction Package website.

**Table 1: Projects general information.**

| | projects | | median age | | contributors | | commits | | pull requests | | coverage builds | | CI builds | |
|---|---|---|---|---|---|---|---|---|---|---|---|---|---|---|
| **size** | **ML** | **nonML** | **ML** | **nonML** | **ML** | **nonML** | **ML** | **nonML** | **ML** | **nonML** | **ML** | **nonML** | **ML** | **nonML** |
| small | 8 | 14 | 5.6 | 7.5 | 28 | 73 | 5497 | 6103 | 2425 | 4511 | 811 | 1054 | 3200 | 4049 |
| medium | 44 | 38 | 5.4 | 9 | 69 | 368 | 37081 | 34545 | 20900 | 31766 | 6736 | 7647 | 35678 | 32983 |
| large | 41 | 40 | 6.6 | 8.7 | 187 | 624 | 100700 | 128988 | 110704 | 84506 | 28770 | 30086 | 161577 | 111992 |
| **total** | 93 | 92 | 5.7 | 8.4 | 102 | 387 | 143278 | 169636 | 134029 | 120783 | 36317 | 38787 | 200455 | 149024 |

***Approach.*** To address RQ1, we use a dataset comprised of 93 ML projects and 92 non-ML projects, as detailed in Section 2.1. First, we compute the median for each of the four CI practices per project, as the data for each project is provided in monthly intervals. However, not all projects have test coverage data available in the CodeCov or Coveralls platforms. For the test coverage metric, we used data from 59 projects (33 ML and 26 non-ML). To assess the statistical differences in the investigated CI practices between ML and non-ML projects, we used Mann-Whitney-Wilcoxon (MWW) tests [29], followed by Cliff's delta effect-size measurements [30]. This analysis was performed while considering project sizes (small, medium, large). The MWW test is a non-parametric test with a null hypothesis assuming that two distributions come from the same population ($\alpha = 0.05$). Cliff's delta is a non-parametric effect-size metric to quantify the magnitude of differences between the values of two distributions. A higher Cliff's delta value indicates a greater disparity between distributions. In interpreting Cliff's delta, we use the thresholds outlined by Romano et al. [31]: $delta < 0.147$ (negligible), $delta < 0.33$ (small), $delta < 0.474$ (medium), and $delta >= 0.474$ (large). Positive Cliff's delta values show how larger are the values of the first distribution, while a negative Cliff's delta indicates the opposite.

## RQ2. What are the evolution trends of build duration and test coverage within ML and non-ML projects?

***Motivation.*** The results of RQ1 show that ML projects tend to require longer build duration compared to non-ML projects and have less test coverage in the case of medium-sized projects. As such, in RQ2, we analyze how the build duration and test coverage of ML and non-ML projects evolve over time. This analysis allows us to understand better the specificities of the differences observed in RQ1. For example, do ML projects have an increasing trend of longer build duration? Or has it always been the case since the beginning of the ML projects?

***Approach.*** To address RQ2, we use the same dataset of 185 projects (93 ML and 92 non-ML) used in RQ1. We compare the ML projects against the non-ML projects to study the potential differences in the build duration and test coverage trends. Both the evolution of build duration and test coverage over time can be interpreted as time series data. Therefore, we use the Dynamic Time Warping (DTW) technique to measure the similarity between different time series, allowing for effective clustering of projects' time series.

The DTW algorithm [32] is a robust method used to measure the similarity between time series by aligning their offsets [33, 34]. This alignment allows for grouping similar time series even if they span different time periods. Before applying the clustering algorithm, it is necessary to determine the optimal number of clusters to obtain high-quality clusters (i.e., clusters that better represent the trends) [35]. To achieve it, we use the gap statistics [36] by varying the number of clusters from 2 to n-1, where n is the number of projects per category. We use the `clusgap` algorithm from `cluster R` package [37] to calculate the gap statistics. After grouping the projects' time series into clusters, we qualitatively analyzed the cluster's centroid to determine their trend, as employed by previous work [17, 35], therefore we don't use specific slope boundaries. These trends were classified as *Increasing*, *Decreasing*, or *Maintaining*. Lastly, we conducted a thorough analysis to determine the proportions of small, medium, and large ML and non-ML projects associated with each identified trend. This comprehensive approach allowed us to discern nuanced patterns within the build duration trends, shedding light on the complex nature of these projects' development processes.

## RQ3. What do ML and non-ML developers discuss about CI in their projects?

***Motivation.*** Considering the notable differences observed between ML and non-ML projects concerning CI in RQ1 and RQ2, we delve deeper in RQ3 to better understand the discussions about CI in both ML and non-ML projects. Our goal is to investigate whether ML and non-ML projects differ in their discussions concerning the use of CI. For instance, ML projects may discuss more build issues, as we observed that they typically require a longer build duration (see RQ1).

***Approach.*** To investigate RQ3, we adopted the qualitative method of document analysis [38, 39]. This approach consists of coding document content into themes. One of the key advantages of document analysis is that documents are "non-reactive". This implies that researchers can read and revisit documents multiple times without being changed by the research process. Our focus is on discussions about CI among developers in both ML and non-ML projects. As pull requests (PRs) encapsulate developers' discussions related to committed code, project infrastructure, and practices, we consider the whole sequence of conversations (i.e., comments) associated with a PR and the PR itself as a document, which is used as the input for our document analysis. Additionally, given that CI workflows are typically triggered after a push or PR submission in open-source projects [40], the CI result for a PR may yield valuable discussions regarding CI among developers. To identify these CI-related conversations, we used a regular expression (Listing 1) to detect CI-related terms in PR comments. Additionally, we aim to exclude bot-generated comments from our analysis by filtering out comments from GitHub users with known bot logins (e.g., coveralls) or logins suggestive of bot users (e.g., vue-bot). Listing 2 represents the regular expression employed to eliminate bot comments from our analysis.

**Listing 1: Regex to detect CI-related terms in PR comments.**

```
\b(ci|continuous integration|devops|dev ops|
workflow(s)?|build(?:s|ing)?|pipeline(s)?)\b
```

**Listing 2: Regex to detect PRs comments from bot users.**

```
\b(bot|ci|codecov|coveralls|github-actions|
dependabot|bors|seldondev|appleboy|netlify|
vercel|azure|pipeline(s)?)\b
```

We identified 11,549 pull requests (7,062 from ML projects and 4,487 from non-ML projects) created post-GitHub Actions adoption, containing CI-related comments from non-bot users across our 185 examined projects. For a balanced representation and to make our manual analysis feasible, we generated two representative samples—365 PR from ML projects and 354 from non-ML projects—employing a Stratified Random Sample (SRS)[41] strategy with a 95% confidence level and a 5% confidence interval. After calculating the sample size, we considered the variability of PRs per project. Therefore, our stratification criteria aimed to maintain the proportion of PRs per project within our population.

To analyze the 719 PRs of our stratified sample, we employ an inductive thematic analysis, which is designed for identifying, analyzing, and reporting themes within qualitative data [42], following the guidelines by Nowell et al [43]. The first step in the thematic analysis is the coding of our data. This step involves attaching codes to any piece of relevant qualitative data collected (i.e., CI discussion) from the document. The first author conducted open coding sessions for all 719 PRs, the second author independently coded 365 PRs from ML projects, and the third author coded 354 PRs from non-ML projects. All PRs were coded by two authors, mitigating bias. In conjunction with the first author, the fourth author collated all codes into a consistent set of refined codes, which often involved adding or merging the codes generated independently by the other authors. The first author then organized these refined codes into themes through axial coding, a process double-checked by the second author. We adopted a reflexive thematic analysis [44], therefore we do not prioritize *coding reliability* at this stage (i.e., we did not calculate the inter-rater agreement between the annotators). This choice aimed at generating themes in developer discussions without the constraint of rigid agreement metrics. It is important to note that this analysis aims to broaden our comprehension of how developers discuss CI in ML and non-ML projects. Our focus is on generating qualitative results that contribute to a deeper understanding of CI discussions in both ML and non-ML projects. The results of this thematic analysis, including codes and themes, are presented in the results section.

# 3 RESULTS

## RQ1. To what extent do CI practices adoption differ between ML and non-ML projects?

***ML projects consistently require longer build durations compared to non-ML projects.*** Figure 4 shows the build duration distributions per project category (ML and non-ML) and project-size category (small, medium, and large). The MWW test and Cliff's delta results are also shown in Figure 4.

Small ML projects require a significantly longer median build duration (10.3 minutes) compared to small non-ML projects (0.813 minutes). The MWW test indicates a significant difference ($p$ = 0.00298, with a *large* Cliff's delta of 0.786). Regarding medium-sized projects, ML projects require a median of 13 minutes for builds, while non-ML projects typically require a median of 5.54 minutes. Our tests indicate a significant difference ($p = 0.000566$, with a *medium* Cliff's delta = 0.437). As for large-sized projects, ML projects require a median build duration of 21.4 minutes, compared to 17.4 minutes in non-ML projects. However, our tests reveal that this difference is not statistically significant ($p = 0.24$). Overall, our findings suggest that ML projects tend to require longer build durations, especially in small-sized and medium-sized projects.

We then noticed that more than half (48 out of 93) of the ML projects have Python, a dynamically typed language, as their main language, while only 14% (13 out of 92) of the non-ML projects have Python as their main language. Overall, the majority (≈ 66%) of the ML projects are based on dynamically typed languages (e.g. Python, JavaScript), while most non-ML projects (≈ 54%) use statically typed languages (e.g., TypeSript, Go). As static typing allows a compiler to perform more optimizations, programs written in statically typed languages tend to be faster than those written in dynamically typed ones [45].[4] Thus, we conjectured that ML projects may require longer build durations than non-ML ones because ML projects are mostly written in dynamically typed languages, mainly in Python, and this would imply in more time for test execution.

To confirm this hypothesis, we compared the building duration of projects based on their programming language type (static or dynamic). For example, we compared the building duration of medium-sized ML projects based on static-typed languages with medium-sized ML projects based on dynamic-typed languages. Contrary to our expectation, we did not find a significant statistical difference (i.e., MWW test $p - value <= 0.05$) in the building duration of ML and non-ML projects of similar size due to the language type system used in the project.

We then performed a second analysis, in which we focused on the building duration difference between similar-sized ML and non-ML projects based on static and dynamic typed languages. This analysis specifically targeted medium-sized ML and non-ML projects. We used the MWW test to explore potential differences in build duration of medium-sized statically typed ML projects with statically typed non-ML projects, as well as medium-sized dynamically typed ML projects with dynamically typed non-ML projects. Small-sized projects were intentionally excluded from this analysis because of the limited group size resulting from categorizing small-sized projects based on their programming language type. This categorization resulted in 4 statically typed and 5 dynamically typed ML projects, along with 5 statically typed and 9 dynamically typed non-ML projects. Furthermore, large-sized ML and non-ML projects were also excluded from the analysis, as no statistically significant difference in build duration was identified between these groups.

[4] https://benchmarksgame-team.pages.debian.net/benchmarksgame/index.html

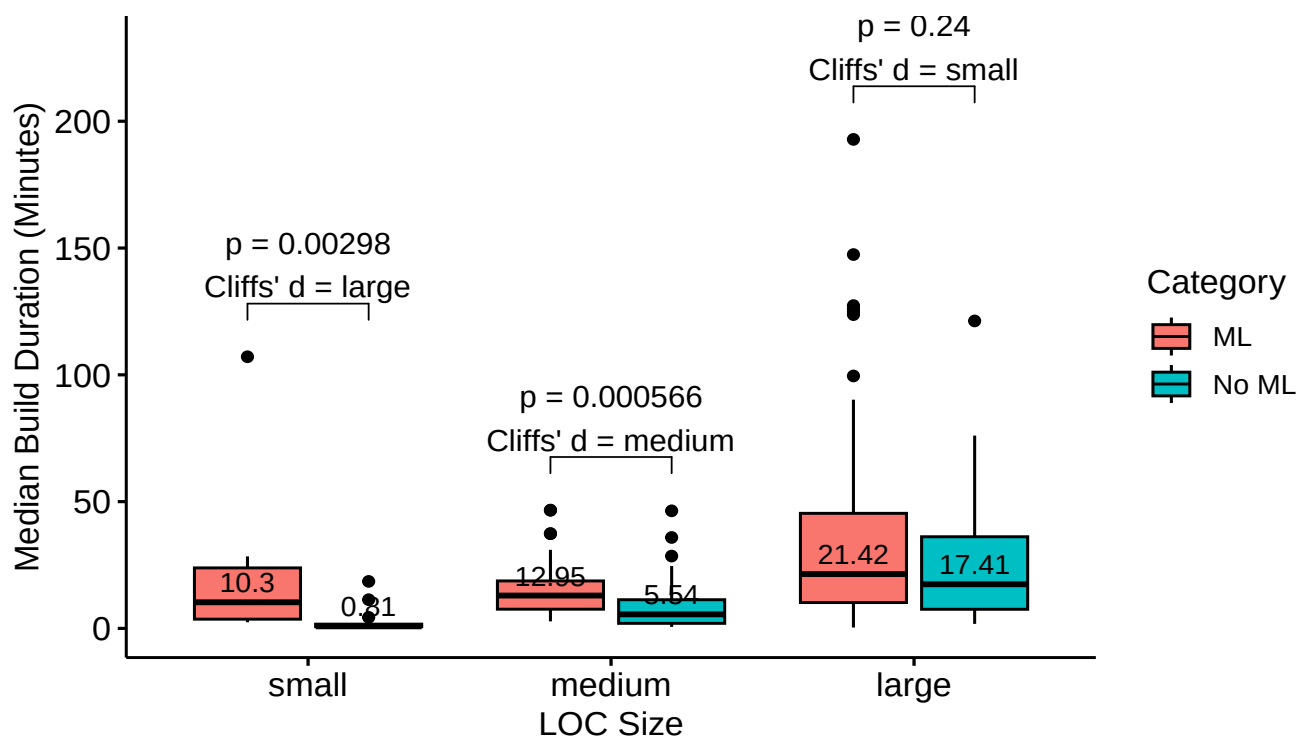

Figure 4: Build Duration per project category and size.

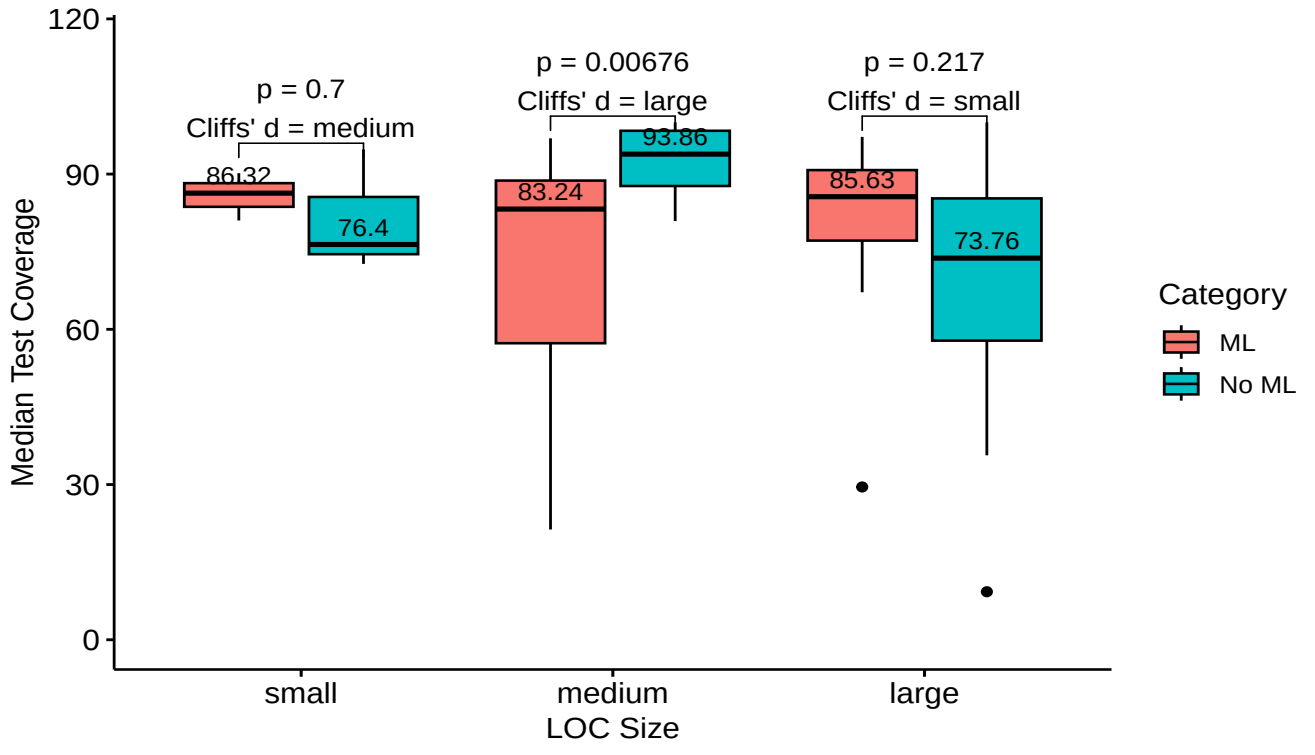

Figure 5: Test Coverage per project category and size.

We found statistically significant differences both between medium-sized dynamically typed ML and non-ML projects (p-value = 0.013, Cliffs' delta = 0.421 (medium)), and between medium-sized statically typed ML and non-ML projects (p-value = 0.044, Cliffs' delta = 0.445 (medium)). This variance in build duration, while controlling for language type, suggests that there is no conclusive evidence to support the initial hypothesis that dynamically typed languages in ML projects lead to longer build duration compared to non-ML projects. Therefore, it implies that other factors may contribute to the extended build duration of ML projects, such as the computationally intensive nature of data pre-processing. The plots illustrating the build duration of medium-sized projects per programming language type are accessible in our online appendix.

***Medium-sized ML projects tend to have a lower test coverage (83%) compared to non-ML counterparts (94%).*** Medium-sized non-ML projects have a substantially higher test coverage compared to their ML counterparts. Figure 5 shows the distributions of test coverage per project category and size. Although large-sized ML projects show higher test coverage levels compared to large non-ML projects, we observe that this difference is statistically insignificant ($p = 0.217$).

Moreover, we conducted a comparison of the metrics "time to fix broken builds" and "commit activity" between ML and non-ML projects. However, no statistically significant differences were identified.

## RQ2. What are the evolution trends of build duration and test coverage within ML and non-ML projects?

As we observed statistically significant differences when it comes to build durations and test coverage in RQ1, we investigate in RQ2 potential differences in how these CI practices evolve over time across ML and non-ML projects.

Upon applying the `clusgap` algorithm to the build duration and test coverage over time (in both ML and non-ML projects), we observe distinct optimal cluster numbers. Concerning the *build duration* metric, the `clusgap` algorithm suggests 10 distinct trends in ML projects, whereas 7 distinct trends emerged in non-ML projects. Concerning *test coverage*, 2 trends emerged in ML projects and 5 trends emerged in non-ML projects.

Figures 6 and 7 illustrate the distinct trends of build duration for ML and non-ML projects. Our focus is on clustering projects with similar build duration trends rather than observing the raw build duration values. To enable a meaningful comparison, build duration was normalized for each project, given that projects of different sizes tend to require varied build durations (i.e., smaller projects typically require shorter build durations than larger ones). Each cluster's trend is represented by a red line indicating a regression linear model calculated using its centroid, facilitating manual classification of trends. We classify each cluster trend as *increasing*, *decreasing*, or *maintaining* for the purpose of interpreting the results. After clustering build duration trends, we categorized the projects based on size and trend type. Figure 8 shows the ratios of small, medium, and large projects that have increasing, decreasing, or maintaining build duration trends.

***Small and medium-sized ML projects exhibit more increasing build duration trends compared to non-ML projects.*** Our results reveal a consistent higher proportion of increasing build duration trends among small and medium-sized ML projects. In smaller ML projects, a significant 75% manifest rising build duration trends. Similarly, 61.4% of medium-sized ML projects display increasing build duration trends. Conversely, small and medium non-ML projects display relatively lower percentages of increasing build duration trends, standing at 35.7% and 44.7% respectively. Although these projects demonstrate relative success in maintaining or decreasing build duration trends over time, there is still room for improvement in their build processes.

***Large ML projects represent a lower proportion of projects displaying an increasing trend (51.2%) compared to non-ML projects (65%).*** The proportion of projects manifesting increasing build duration trends diminishes as ML projects expand in size (small: 75%, medium: 61.4%, large: 51.2%). In contrast, non-ML projects exhibit an opposite tendency, with increasing proportions

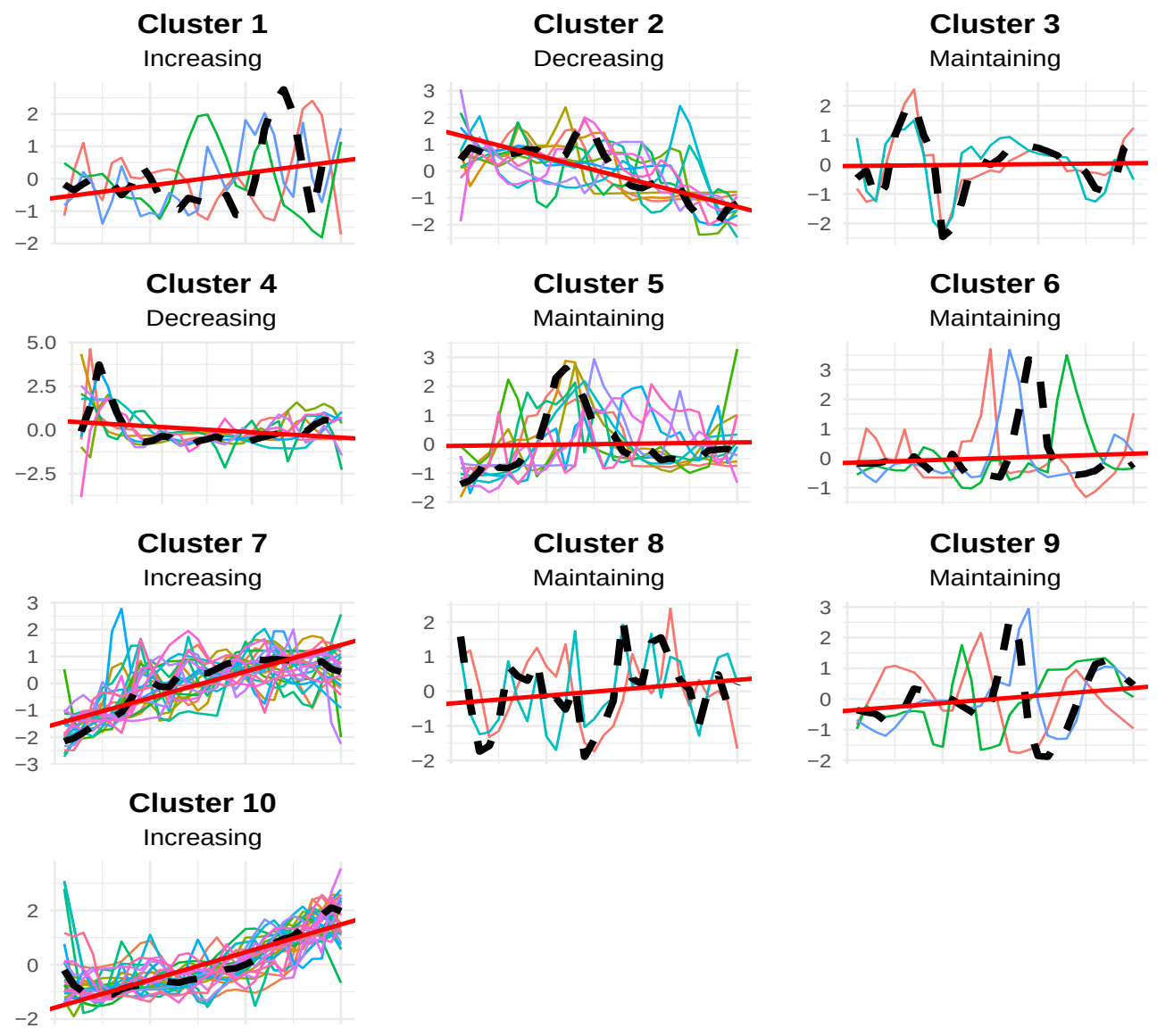

**Figure 6: Build Duration Clustering Trends' Patterns in ML Projects.**

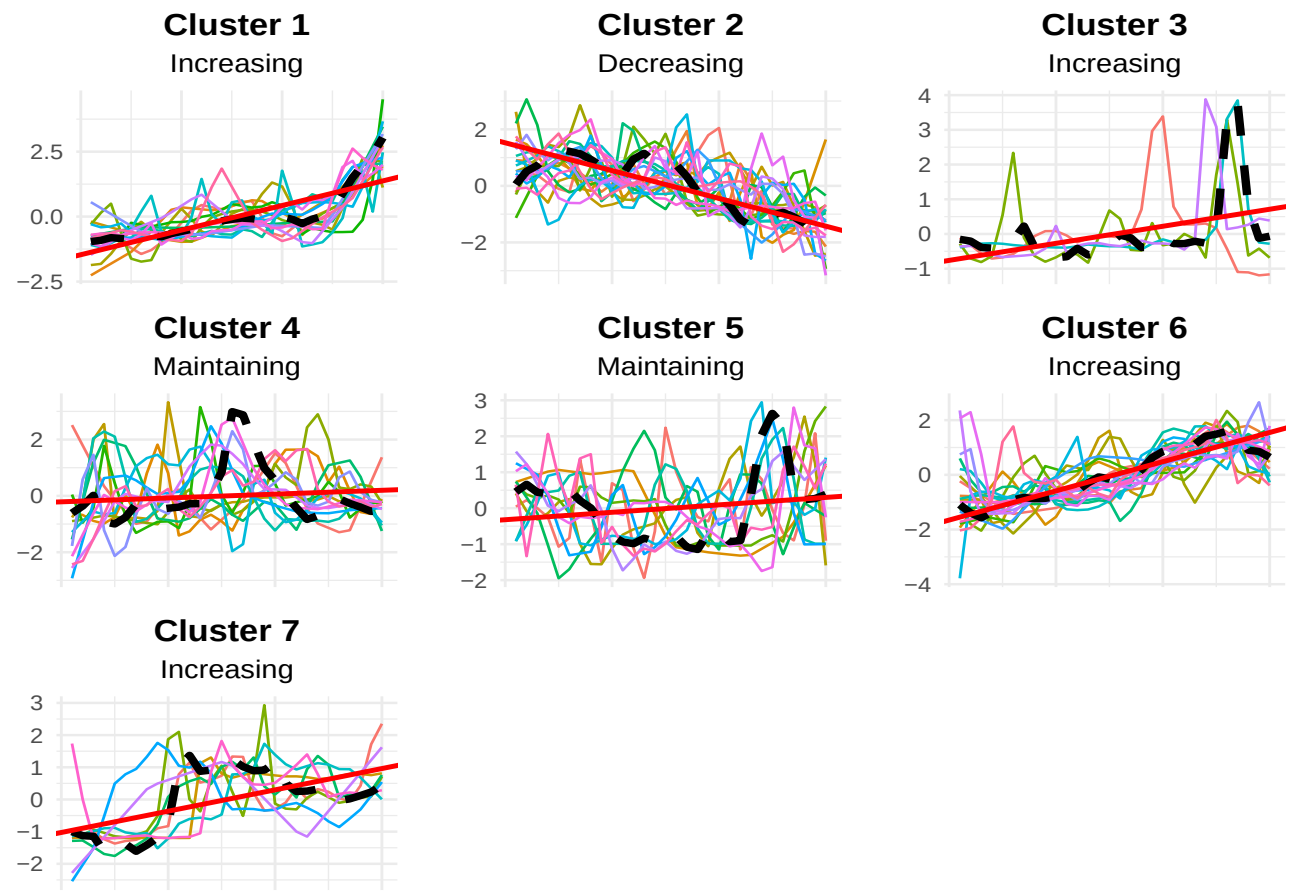

**Figure 7: Build Duration Clustering Trends' Patterns in non-ML Projects.**

observed as project sizes grow (small: 35.7%, medium: 44.7%, large: 65%). Nevertheless, both types large-sized projects represent substantial proportions of rising build duration trends.

***Both ML and non-ML projects manifest a predominant trend of maintaining test coverage over time.*** While we found a statistically significant difference in test coverage between medium-sized ML and medium-sized non-ML projects in *RQ*1, no noticeable differences in test coverage trends were observed between these project types. Specifically, medium-sized non-ML projects consistently manifest a maintaining test coverage trend, with all projects with a median coverage rate above 75%. In contrast, 46.6% (7/15) of the ML projects manifest a test coverage rate lower than 75%, with two of them exhibiting median coverage rates of 21% and 34%. Moreover, these ML projects with low test coverage rates also lack rising trends of test coverage. The plots that represent the cluster trends for test coverage are available in our online appendix.

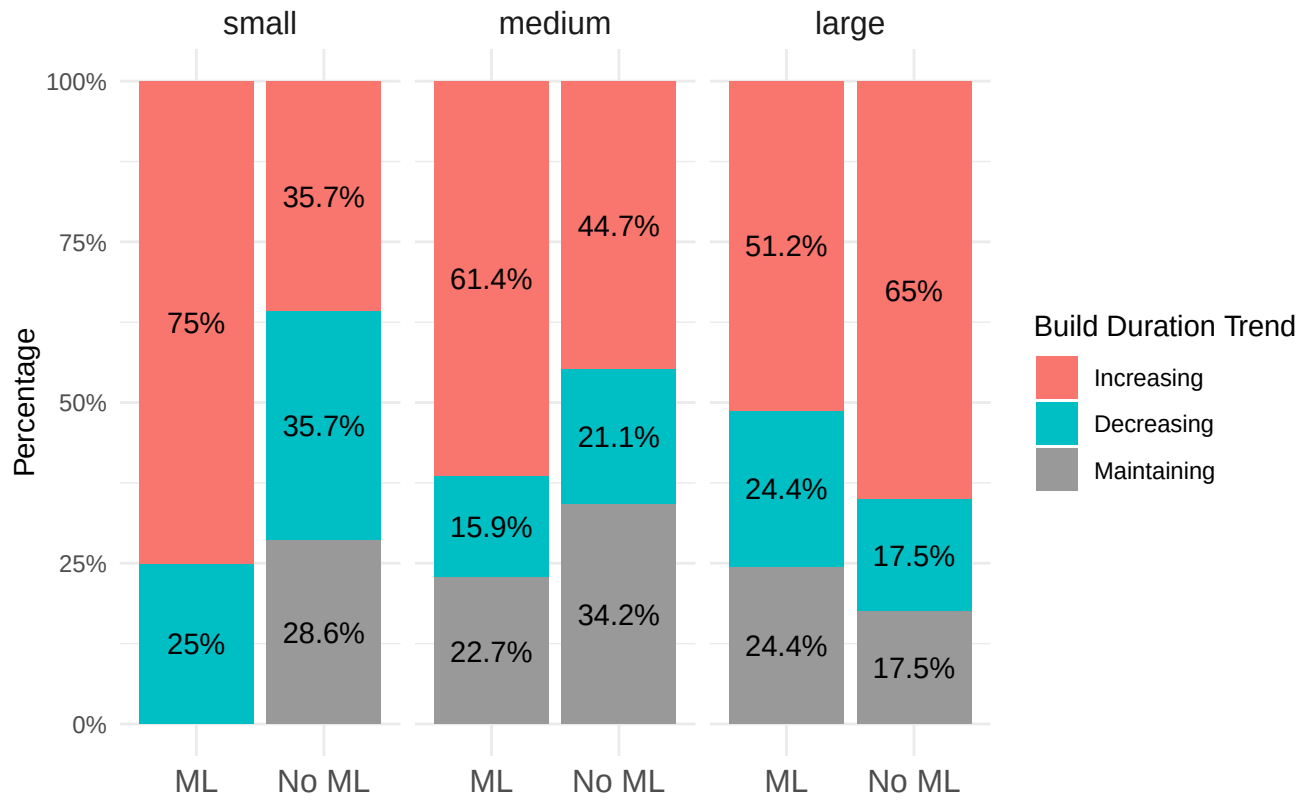

**Figure 8: Clustering trends per project category and size.**

## RQ3. What do ML and non-ML developers discuss about CI in their projects?

Figure 9 presents the CI discussions (i.e., the themes) that emerged from our document analysis. The central theme is *CI discussion*, which is the main objective of this RQ. The second-level (or axial) themes comprise discussions related to CI: *CI build execution and status*; *CI testing and code quality*; *CI infrastructure*; *CI pipeline configuration*; and *CI on software development process*. The third-level themes are more specific and are grouped based on their relationship with the second-level themes. The thickness of the edges is based on the number of times the third-level theme emerged during the thematic analysis. We provide in-depth details about each theme that emerged in our thematic analysis in our online appendix.

In the first exploratory analysis, we observed similar engagement levels in developer comments on CI discussions of ML and non-ML projects (ML: median of 6 comments; non-ML: median of 7 comments) and in the number of codes associated with CI discussions (ML: median of 2 codes, non-ML: median of 2 codes). Both non-ML and ML projects commonly discuss themes related to *CI Build Execution and Status*, *CI Infrastructure*, *CI Pipeline Configuration*, and *CI Testing and Code Quality*. Despite these similarities, we found a wider variety of CI-related themes in ML projects (73 themes) compared to non-ML projects (23 themes).

Interestingly, discussions on the impact of CI on the software development process were exclusive to ML projects, including themes such as *(mis)trust in CI*, *CI as a decision-making* tool, and *CI as a quality gate*. Notably, comments such as *"going to hold off on review until the tests run successfully"* were coded as CI acting as a quality gate. Additionally, the *(mis)trust in CI* code, which emerged only for ML projects, indicates the developers' uncertainty about the effectiveness of their CI system in catching project-specific bugs: *"I don't think our CI will catch all such bugs"*.

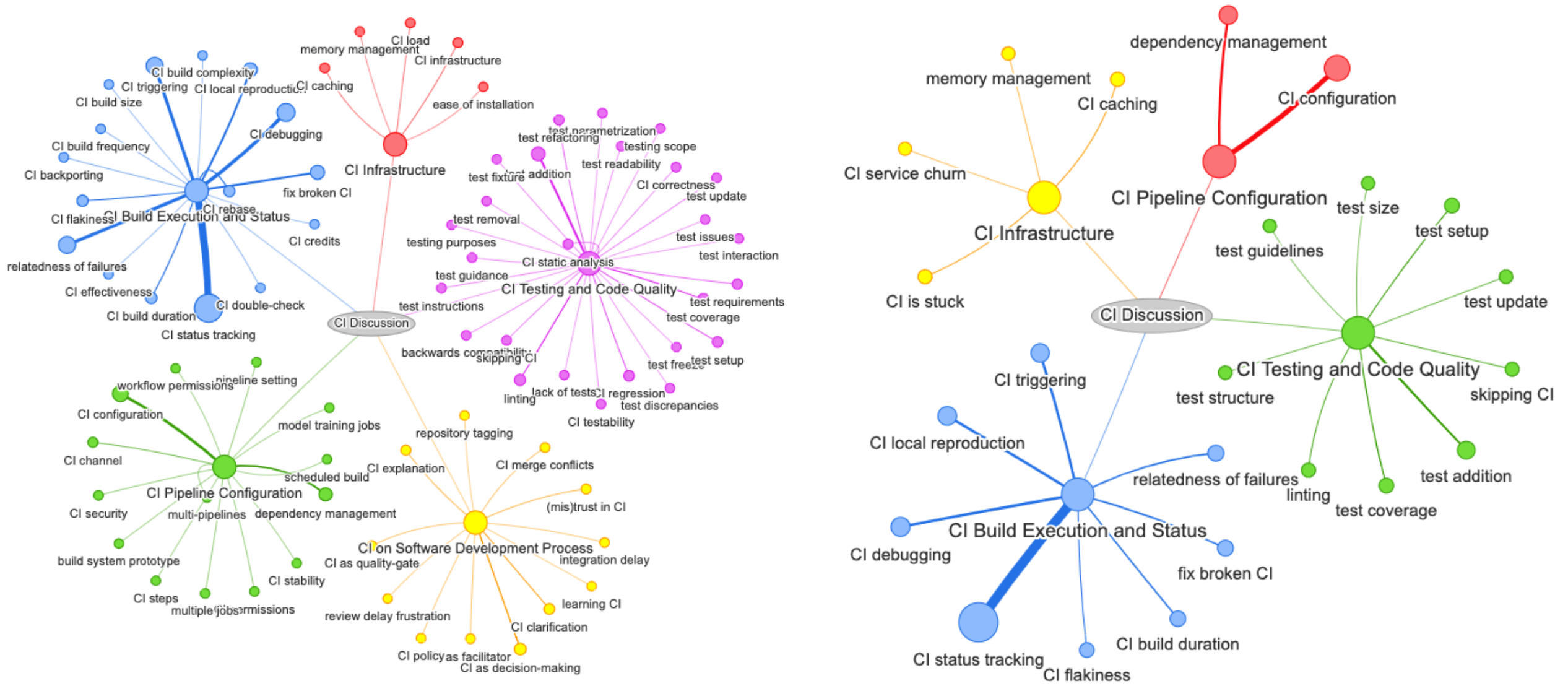

**(a) CI discussions in ML projects.**

**(b) CI discussions in non-ML projects.**

**Figure 9: CI discussions in ML and non-ML projects.**

Our analysis also highlights a significant convergence in key CI discussion themes in both ML and non-ML projects, with topics like *CI status tracking*, *CI debugging*, *CI triggering*, *CI configuration*, *fixing broken CI*, *CI local reproduction*, and *test addition*. This emphasizes a shared commitment to maintaining code quality through robust testing practices across diverse software development domains. However, a notable distinction arises in the prevalence of the theme *relatedness of failures*, particularly in ML projects, where it ranks as the third most discussed theme. In contrast, it holds the eighth position in discussions within the non-ML domain. This theme sheds light on concerns about potential false positives in CI systems, where although a build is broken, developers deem that such failure is not related to their current push. These false positives not only introduce frustration but also result in the expenditure of valuable time and effort by developers and maintainers in addressing issues unrelated to their code modifications. For instance, a developer of the *scikit-learn* project expressed frustration about build breakages unrelated to a PR, stating: *"I would not want PRs to have failing tests because of a poor random seed that is unrelated to the PR itself. That feels like a poor contributor experience"*. This finding aligns with our earlier discovery about the theme of *(mis)trust in CI*, which exclusively emerged in ML projects. Both themes suggest a unique set of challenges in ML projects regarding the effectiveness and reliability of their CI systems, emphasizing the need for a more nuanced and tailored approach to CI practices in this specific domain.

# 4 DISCUSSION AND IMPLICATIONS

**Interpreting differences in CI practices between ML and non-ML projects.** In RQ1, we observed that ML projects consistently require longer build duration compared to non-ML projects. Notably, a significant portion of ML projects (48 out of 93) primarily use Python, a dynamically typed language, in contrast to non-ML projects where only 14% (13 out of 92) use Python. The prevalence of dynamically typed languages in ML, known for potential runtime flexibility but longer running times, may be a contributing factor to the observed longer build duration. Moreover, possible factors contributing to this disparity include the computationally intensive nature of data preprocessing in ML projects, which adds overhead to the build process. Challenges in parallelization during the build, especially in workflows with large datasets, may further contribute to extended build times. The diverse testing requirements of ML, involving numerous model configurations and data inputs, can also lead to longer builds due to the need for extensive validation. In future research, we intend to study in more depth the impact of these factors on the build duration of ML projects.

Additionally, we found that medium-sized ML projects tend to exhibit lower test coverage (83%) compared to their non-ML counterparts (94%). The testing landscape in ML projects often involves data-driven testing, where the diversity and complexity of data scenarios present challenges in achieving comprehensive test coverage [46]. Generating representative datasets for testing all possible scenarios can be more demanding in ML projects. Moreover, ML models frequently employ different algorithms, each with unique characteristics, adding complexity to achieving comprehensive test coverage across diverse algorithms and their combinations. Additionally, ML systems consist of multiple interconnected components, such as data preprocessing, model training, and deployment [47]. In addition to that, recent research is exploring new ways of coverage for ML systems [46]. Achieving high test coverage across the entire integrated system becomes challenging due to the intricacies of testing interactions between these components and the nature of ML systems that maintain decision logic learned from training data.

Furthermore, the observed similarity in "time to fix broken builds" and "commit activity" metrics across both ML and non-ML projects suggests a degree of uniformity in certain development challenges. The finding implies that, despite the distinct nature of ML projects, developers do not exhibit significantly different response times in resolving build issues for ML projects compared to non-ML projects. Consequently, the data suggests that build issues do not inherently pose a more severe challenge in ML projects. This insight may guide practitioners in allocating resources and prioritizing efforts to address development challenges that are more distinctive to the specific characteristics of ML projects (i.e., build duration).

Intriguingly, in RQ2, we noted a distinctive trend in the build duration of ML and non-ML projects: while non-ML projects tend to exhibit an increasing trend with project size (small, medium, large), ML projects show a divergent pattern as they expand. Considering that ML projects require longer build durations compared to non-ML projects (see RQ1), with build duration taking more than 10 minutes even for small projects, it might be the case that project maintainers of ML projects could lead to continuous efforts (i.e., workflow parallelization, caching) to enhance efficiency and control increasing trends as the project grow.

**Interpreting developers' discussions about CI.** In our qualitative analysis (RQ3), we find that both ML and non-ML projects commonly discuss themes related to *CI Build Execution and Status*, *CI Infrastructure*, *CI Pipeline Configuration*, and *CI Testing and Code Quality*, which demonstrate a commitment to effective CI practices. However, we found a wider variety of CI-related themes in ML projects (73 themes) compared to non-ML projects (23 themes). Several factors may contribute to this diversity in CI-related discussions within ML projects. For instance, ML projects might attract developers with other backgrounds than software engineering, including data scientists and ML specialists. This diversity in expertise could lead to a more comprehensive discussion of CI practices tailored to the specific needs of ML development. Additionally, the intricate workflows in ML, involving stages such as data preprocessing, and model training and deployment, may lead to a more extensive range of CI-related discussions addressing diverse challenges at each stage. Hence, a future study may investigate how the varied development backgrounds of ML project developers impact the depth of CI discussions in their projects. Specifically, exploring whether developers with diverse skills or more CI experience contribute to the richness of CI-related discussions in ML projects.

Additionally, a notable disparity emerges in the prevalence of the "relatedness of failures" theme within CI discussions of ML projects. The prominence of this theme in ML projects suggests a potentially higher incidence of false positives in their CI systems. Moreover, false positives might contribute to developers' frustration when failing tests are not directly linked to their code modifications, indicating a perceived inefficiency in the CI process. This difference in the occurrence of false positives in CI systems of ML projects may be associated with the intricate workflows and dependencies inherent in ML projects. However, further research is needed to comprehensively understand the root causes of false positives in CI systems, especially within the context of ML projects. Hence, a future study may investigate how the development backgrounds of developers of ML projects impact the depth of CI discussions in their projects.

**Implications for practice and research.** Our results have implications for practice. Firstly, the observation of longer build duration in ML projects compared to non-ML projects underscores the importance of addressing unique challenges posed by ML workflows. Project maintainers of ML projects should focus on continuous efforts, such as workflow parallelization and caching, to enhance efficiency and control the increasing build duration trends, especially as the projects grow in size. This highlights the importance of proactive optimization strategies in ML project development processes to mitigate the challenges associated with longer build duration [48]. Furthermore, the prevalent use of dynamically typed languages in ML projects, particularly Python, should prompt project maintainers to explore strategies for minimizing compilation times.

The lower test coverage observed in medium-sized ML projects compared to non-ML counterparts emphasizes the need for tailored testing approaches in the ML domain. Practitioners should recognize the intricacies of data-driven testing in ML projects, where diverse and complex data scenarios pose challenges for achieving comprehensive test coverage.

Furthermore, acknowledging the prominence of discussions on false positives in CI outcomes within ML projects, project maintainers—particularly those in the ML domain—can proactively assess whether their projects' CI systems are susceptible to this issue. This heightened awareness enables practitioners to implement targeted strategies for mitigating false positives, thereby fostering a more reliable and efficient CI environment.

## 5 THREATS TO VALIDITY

**Construct validity.** We define *CI Workflows* as GitHub Actions workflows that run a CI pipeline. To identify the workflows that represent *CI workflows*, we looked for actions that contain CI-related words (such as 'test' or 'continuous integration') in the *job* or step *name* of their configuration file. However, *CI workflows* may not match this criterion but execute a CI pipeline (i.e., build and run a test suite of a system). We manually inspected the configuration files of a random sample of 100 GitHub Actions workflows and compared the sample with the criterion's result. Our selected criterion achieved 86.7% recall and 100% precision.

**Internal validity.** We started by using a dataset containing ML and non-ML GitHub projects provided by the work of Gonzalez et al. [4], which acknowledges that their project selection criteria may lead to false positives and negatives. Rzig et al. [9] revised Gonzalez et al. [4] dataset, confirming that their projects are real projects involving ML, filtering out toy projects and tutorials. The projects selected for our study were extracted from the Rzig et al. [9] dataset which provides a more curated list of ML and non-ML projects that adopted CI practices. Additionally, for gathering coverage information, we used the Coveralls and Codecov APIs, and for obtaining the number of lines of code in the studied projects, we relied on the CodeTabs API. It is important to notice that, as a proprietary third-party service, we acknowledge that we had to depend on its output without direct control over the data collection process.

Moreover, we recognize that the conclusions drawn from our analysis depend on project selection. To ensure fairness, we carefully curated comparable sets of ML and non-ML projects. Our study includes diverse ML projects (tools and applications) and non-ML projects (e.g., Static Site Generators, Cloud Tools, Web/Mobile Frameworks, package management, etc). As such, we consider that the sets of projects are comparable in terms of complexity. Additionally, we tried to minimize confounding effects by selecting similar-sized projects in both categories. We can see in Table 1 that for some metrics (e.g., CI builds, commits), the numeric difference between both groups is small (less than 10%), while for others (e.g., median age), we have a bigger discrepancy. This may have affected the results, but we consider that their impact does not invalidate our findings.

In the qualitative analysis of RQ3, we recognize the potential influence of subjective interpretations by the authors regarding PR comments related to CI discussions. To mitigate this threat, we implemented a rigorous process involving peer review of the codes and themes extracted through document analysis. Two distinct authors independently coded each document (i.e., PR and its comments) in our analysis. Furthermore, a third author consolidated the diverse codes into a new set, resolving any discrepancies that emerged between the independently generated codes. This approach enhances the robustness and reliability of our qualitative findings.

**External validity.** We focus our analysis on 185 public repositories from GitHub. All projects were set up to use GitHub Actions as their CI service. We acknowledge that we cannot generalize our results to other coding hosting platforms (i.e., GitLab), private GitHub projects, projects in companies, or those using different CI services. However, we believe our study can provide a good understanding of how CI is adopted in both ML and non-ML projects, as we study a varied set of projects with different programming languages and sizes.

## 6 RELATED WORK

Prior research has investigated the impact of CI in the context of open-source projects [13–17]. For example, Vasilescu et al. [13] conducted a study with over 200 GitHub projects using Java, Ruby, or Python as the main programming language, revealing a positive association between the adoption of Travis CI and the number of bugs detected by core developers. Hilton et al. [16] observed that CI plays a pivotal role in facilitating more frequent software releases. Furthermore, Bernardo et al. [15] found that after adopting Travis CI, projects deliver a higher proportion of PRs per release. Different from these works, which usually investigate the dynamics of software development projects after adopting a CI service (i.e., Travis CI), our study investigates how specific CI practices are employed in the context of ML projects.

Recent works on CI have expanded their focus beyond merely adopting CI services, instead investigating the specific CI practices employed by software projects [18, 19]. Felidré et al. [18] investigated 1,270 non-ML projects that use Travis CI to understand how these projects face unhealthy CI practices, such as infrequent commits, long time to fix builds, poor test coverage, and long build durations. Their results show that the build is executed under 10 minutes in most of the investigated projects. Similar to the work of Felidré et al. [18], we use the same CI practices investigated in their study to analyze the extent to which the adoption of CI practices differs between ML and non-ML projects. Regarding build duration, our results for non-ML projects align with their findings, unlike ML projects where even small projects have a median build duration longer than 10 minutes.

Specifically in the ML domain, Rzig et al. [49] conducted a study characterizing the usage of CI tools in ML projects. They found that in a large dataset of popular GitHub projects (4031 ML and 4076 non-ML projects), about 37% of ML projects and 45% of non-ML projects have adopted a CI service. Our work uses the dataset provided by Rzig et al. [49] as the initial dataset used for filtering ML and non-ML projects for our analysis. However, different from their work, which focuses on projects that use Travis CI, our study investigates projects that adopted GitHub Actions, as it emerged as the prevailing CI service on GitHub [50]. Despite the prevalence of CI services like GitHub Actions in ML, recent efforts aim to create dedicated CI services for the field, considering its intricacies (i.e., traditional testing may overfit ML models) [21]. These characteristics introduce variability in the development lifecycle, which may challenge the seamless integration of CI practices. In contrast to prior work, our study specifically compares CI practices between ML and non-ML projects. Also, we perform a clustering analysis to understand the overall trend of metrics such as build duration and test coverage over time. Finally, we uncovered specific themes discussed by developers in both ML and non-ML projects.

## 7 CONCLUSION

In this work, we investigated how CI practices are applied in both ML and non-ML Github projects. We conducted an empirical study that quantitatively analyzed data from 185 open-source projects from Github, comprising 93 ML projects and 92 non-ML projects. Additionally, we conducted a qualitative analysis of a stratified sample of 719 PRs, consisting of 365 ML and 354 non-ML PRs, and their associated comments to shed light on developers' discussions about CI in their projects.

Our study findings highlight distinctive challenges in CI adoption within ML projects, such as longer build duration and reduced code coverage in medium-sized projects. The prevalence of increasing build duration trends in small and medium-sized ML projects underscores the necessity for tailored CI approaches. Moreover, the qualitative analysis uncovers specific themes in CI discussions for both ML and non-ML projects, elucidating the unique challenges faced by ML projects. Overall, the study provides valuable insights into the nuanced landscape of CI practices in ML projects, offering a foundation for customized and effective approaches in this domain.

## ACKNOWLEDGMENTS

This work is partially supported by INES (www.ines.org.br), CNPq grant 465614/2014-0, CAPES grant 88887.136410/2017-00, FACEPE grants APQ-0399-1.03/17, PRONEX APQ/ 0388-1.03/14, and CNPq grant 425211/2018-5.